# AES Implementation and Performance Evaluation on 8-bit Microcontrollers

Hyubgun Lee
Department of Computing
Soongsil University
Seoul, South Korea .

Kyounghwa Lee
Department of Computing
Soongsil University
Seoul, South Korea .

Yongtae Shin
Department of Computing
Soongsil University
Seoul, South Korea
.

*Abstract*— The sensor network is a network technique for the implementation of Ubiquitous computing environment. It is wireless network environment that consists of the many sensors of lightweight and low-power. Though sensor network provides various capabilities, it is unable to ensure the secure authentication between nodes. Eventually it causes the losing reliability of the entire network and many secure problems. Therefore, encryption algorithm for the implementation of reliable sensor network environments is required to the applicable sensor network. In this paper, we proposed the solution of reliable sensor network to analyze the communication efficiency through measuring performance of AES encryption algorithm by plaintext size, and cost of operation per hop according to the network scale.

## I. INTRODUCTION

The sensor network is a network technique for the implementation of Ubiquitous computing environment. It is wireless network environment that consists of the many sensors of lightweight and low-power. It is researching and developing at the various standards and research organizations. As a result, various fields such as logistics, environmental control, home network applied to sensor network [1]. In these environments, the data is collected by sensors is used through the systematic analysis and the cross-linking between services in a variety of services. Therefore, common security requirements (integrity, confidentiality, authentication, non-repudiation) are required for security service and applications.

Public key encryption algorithm is a fundamental and widely using technology around the world. But it has hardware limitations as like memory and battery, so it is not applied to the sensor network [2]. Therefore, Symmetric key encryption algorithm with low-Energy consumption is used in the sensor networks.

In this paper, we describe the Rijndael's AES encryption algorithm in the symmetric key encryption. And we measure the encryption and decryption performance on the 8-bit Microcontroller. Then, we analyse the communication efficiency through the total delay per hop in sensor network.

The structure of the paper is organized as follows: Section 2 describes The Rijndael's AES encryption algorithm in Symmetric key encryption; Section 3 measures the encryption and decryption performance on the 8-bit Microcontroller; Sections 4 analyzes the communication efficiency in sensor network through the total delay per hop; and Section 5 concludes this paper.

## II. AES(ADVANCED ENCRYPTION STANDARD)

### A. *Rijndael's algorithm*

The AES (advanced encryption standard) [3] is an encryption standard as a symmetric block cipher. It was announced by National Institute of Standards and Technology (NIST) as U.S. FIPS PUB 197 (FIPS 197) on November 26, 2001. The central design principle of the AES algorithm is the adoption of symmetry at different platforms and the efficiency of processing. After a 5-year standardization process, the NIST adopted the Rijndael algorithm as the AES.

The AES operates on 128-bit blocks of data. The algorithm can encrypt and decrypt blocks using secret keys. The key size can either be 128 bit, 192 bit, or 256 bit. The actual key size depends on the desired security level. The different versions are most often denoted as AES-128, AES-192 or AES-256.

The cipher Rijndael [4] consists of an initial Round Key addition, Nr-1 Rounds, a final round. Figure 1 shows the pseudo C code of Rijndael algorithm.

```
Rijndael(State,CipherKey) {
    KeyExpansion(CipherKey,ExpandedKey) ;
    AddRoundKey(State,ExpandedKey);
    For(i=1; i<Nr; i++) Round(State,ExpandedKey + Nb*i) ;
    FinalRound(State,ExpandedKey + Nb*Nr);
}
```

Figure 1. Rijndael algorithm

The key expansion can be done on beforehand and Rijndael can be specified in terms of the Expanded Key. The Expanded Key shall always be derived from the Cipher Key and never be specified directly. There are however no restrictions on the selection of the Cipher Key itself. Figure 2 shows the pseudo C code of Rijndael's Expanded Key algorithm.

---

Rijndael(State, ExpandedKey) {

    AddRoundKey(State, ExpandedKey);

    For(i=1; i<Nr; i++ ) Round(State, ExpandedKey + Nb*i);

    FinalRound(State,ExpandedKey + Nb*Nr);

}

---

Figure 2.  Rijndael's Expanded Key algorithm

### B. AES round transformation

The round transformation [5] modifies the 128-bit State. The initial State is the input plaintext and the final State is the output ciphertext. The State is organised as a 4 X 4 matrix of bytes. The round transformation scrambles the bytes of the State either individually, rowwise, or columnwise by applying the functions SubBytes, ShiftRows, MixColumns, and AddRoundKey sequentially. Figure 3 show the AES iterates a round transformation.

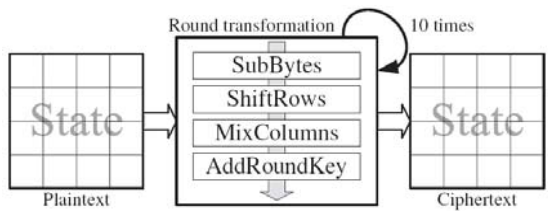

Figure 3.  AES iterates a round transformation.

An initial AddRoundKey operation precedes the first round. The last round differs slightly from the others the MixColumns operation is omitted.

SubByte is a substitution function in the Cipher round. In the SubBytes step, each byte in the state is replaced with its entry using a nonlinear byte substitution table (S-box) that operates on each of the State bytes independently. Figure 4 shows the SubBytes applies the S-box to each byte of the State.

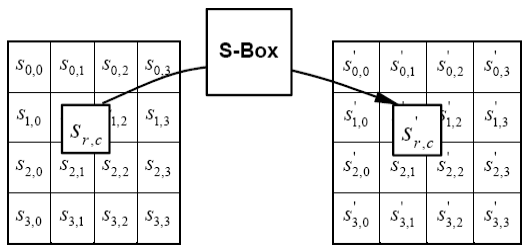

Figure 4.  SubBytes applies the S-box to each byte of the State

ShiftRows is a permutation function in the Cipher round. In the ShiftRows step, bytes in each row of the state are shifted cyclically to the left. The number of places each byte is shifted differs for each row. ShiftRows step is composed of bytes from each column of the input state. Figure 5 show the ShiftRows cyclically shifts the last three rows in the State.

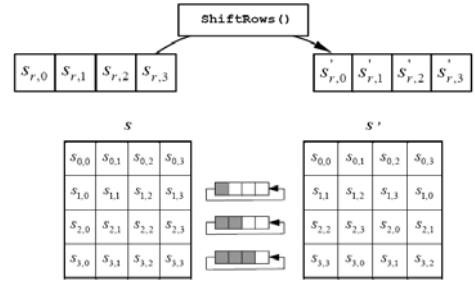

Figure 5.  ShiftRows cyclically shifts the last three rows in the State

MixColumns is a Mixing function in the Cipher round. In the MixColumns step, In the MixColumns step, the four bytes of each column of the state are combined using an invertible linear transformation. The MixColumns function takes four bytes as input and outputs four bytes, where each input byte affects all four output bytes. Together with ShiftRows, MixColumns provides diffusion in the Cipher. Figure 6 shows the MixColumns operates on the State column-by-column.

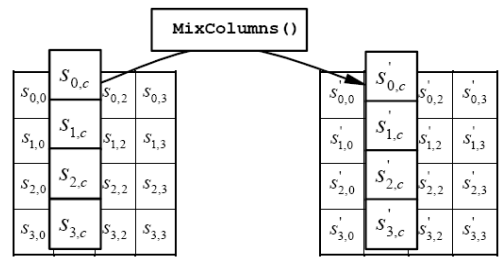

Figure 6.  MixColumns operates on the State column-by-column

AddRoundKey is a key adding function in the Cipher round. In the AddRoundKey step, the subkey is combined with the state. For each round, a subkey is derived from the main key using Rijndael's key schedule, each subkey is the same size as the state. The subkey is added by combining each byte of the state with the corresponding byte of the subkey using bitwise XOR. Figure 7 shows the AddRoundKey XORs each column of the State with a word from the key schedule.

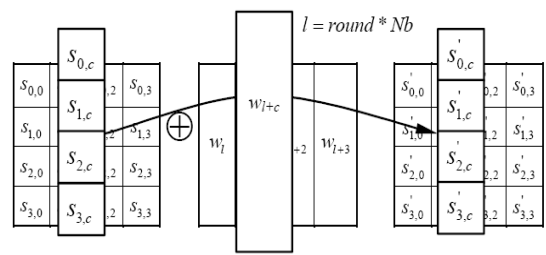

Figure 7.  AddRoundKey XORs each column of the State with a word from the key schedule

AES Decryption computes the original plaintext of an encrypted ciphertext. During the decryption, the AES algorithm reverses encryption by executing inverse round transformations in reverse order. The round transformation of decryption uses the functions AddRoundKey, InvMixColumns, InvShiftRows, and InvSubBytes.

## III. IMPLEMENTATION AND PERFORMANCE EVALUATION

### A. Experiment and Device

For the performance analysis of AES encryption algorithm in the sensor network, we use the ATmega644p [6] in 8-bit Microcontroller as a hardware device. The AVR Studio 4 and Programmer's Notepad in the WinAVR are used as development tools. The JTAG (Joint Test Action Group) Emulator is used as a debugging tool. Figure 8 shows that device for the performance analysis of the AES encryption algorithm.

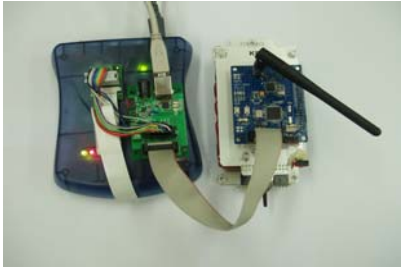

Figure 8. Device for the performance of analysis of AES

ATmega644p in 8-bit Microcontroller is made by Atmel. The main function of the ATmega644p is to ensure correct program execution. It must therefore be able to access memories, perform calculations, control peripherals, and handle interrupts. It has 20Mhz System Clock, prescaler of 8, 64, 256 or 1024 and advanced RISC Architecture.

AVR Studio is execution or debugging without AVR Microcontroller board. And compiled programs are applied to the AVR. Programmer's Notepad with the Win-GCC Compiler compiles the written C language. The compiled programs are applied to the AVR Studio. JTAG Emulator in JTAG Standard is I/O device using JTAG Port which receives the information from PCB or IC.

### B. The implementation of principle

For the performance Measurement of AES encryption algorithm, we apply the AES-128 CBC (Cipher Block Chaining) mode to the ATmega644p's EEPROM. In CBC mode, each block of plaintext is XORed with the previous ciphertext block before being encrypted. Also, to make each message unique, an initialization vector must be used in the first block. Figure 9 shows that CBC mode encryption [7].

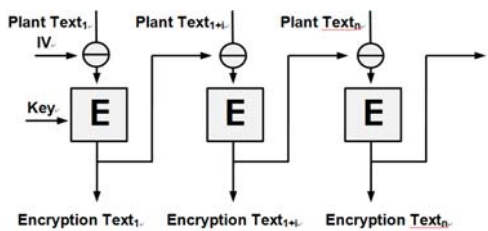

Figure 9. CBC mode encryption

The timer mode for the time measurement uses the Timer/Counter CTC (Clear Timer on Compare Match) Mode. The CTC Mode generates the compare interrupt only if the counter value (TCNT), which is cleared to zero, matches the OCR. The timer measurement measures the counts ($P$) of the compared interrupt per 1ms.

The operation time per 1 clock ($T_C$) is the following:

$$T_C = \frac{1}{Frequency} = \frac{1}{20*10^6} \quad (1)$$

The ATmega644P has a system clock prescaler, and the system clock can be divided by setting the Clock Prescale Register. The prescale time per system clock prescaler ($T_P$) is the following:

$$T_P = prescaler * T_C \quad (2)$$

The Timer/Counter (TCNT) and Output Compare Registers (OCR) are 8-bit Registers. The OCR for the generating of the compare Interrupt is the following:

$$OCR0A = 0xFF - (0xFF - (P/T_P) + 1) \quad (3)$$

### C. Result

For the comparison between encryption and decryption performance, we use the AES-128 CBC mode. The operation time of the encryption and decryption is measured to the data sizes of 16, 32, 64, 128, 256 and 512 Byte. Table 1 and Figure 10 show the encryption and decryption operation time and CPU cycle according to the data size.

TABLE I. THE COMPARISON BETWEEN ENCRYPTION AND DECRYPTION PERFORMANCE BY DATA SIZES

| Data Size(byte) | | 16 | 32 | 64 | 128 | 256 | 512 |
|---|---|---|---|---|---|---|---|
| Enc | Time (ms) | 449 | 898 | 1,796 | 3,592 | 7,184 | 14,368 |
| | CPU Cycle | 8,980 | 17,960 | 35,920 | 71,840 | 143,680 | 287,360 |
| Dec | Time (ms) | 456 | 912 | 1,825 | 3,649 | 7,297 | 14,592 |
| | CPU Cycle | 9,120 | 18,240 | 36,500 | 72,980 | 145,940 | 291,840 |

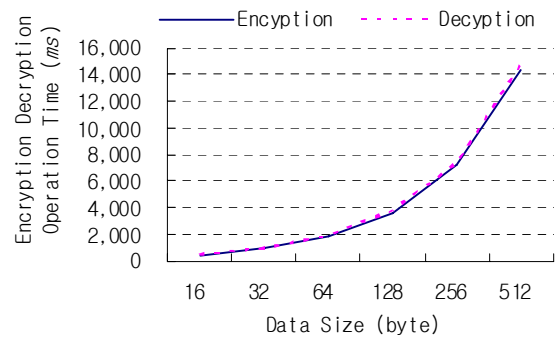

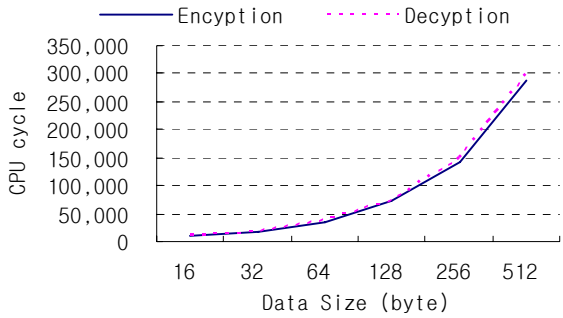

Figure 10. The operation time and CPU Cycle by data sizes

In the result, the operation time and CPU Cycle by data sizes increase approximately 2 times. In 512 byte, it takes approximately 14 minutes to the encryption and decryption.

## IV. APPLICATION SCENARIO

### A. Network model

Figure 11 shows that a general node ($N_i$) sends the secured data packet to the cluster head (*CH*) in the same subnet.

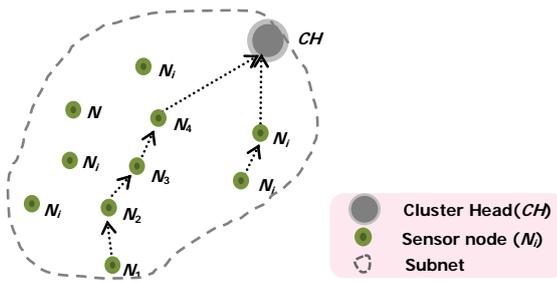

Figure 11. Sensor Network Application Model

For measurement of the data encryption and decryption transmission delay by the number of communication hop, the following assumptions are established. Namely, the every node within subnet has same performance, and there is no interfere or packet loss in the data communication. Each node shares common key with neighbor nodes in advance, and operates encryption and decryption once per hop. The communication for the generating of Pair-wise Shared Key is similar to the μTESLA(Micro Timed Efficient Stream Loss-tolerant Authentication) protocol of the sensor network [8].

### B. Communication delay in sensor network

In communication process of the sensor network, the Beacon Request Command and Association Request Command are communicated between new node and cluster head. The general node ($N_1$) encrypts the data using the pre-deployed security key. It sends secured data to the neighbor node ($N_2$). The node ($N_2$) decrypts the encrypted message ($msg_E$) using the pre-deployed security keys. Then it obtained to the plantext. The node ($N_2$) repeats the same process in the previous step using the private key shared with its neighbor node ($N_3$).

The data delivery process by hop communication is following:

$$\forall N_i \in subnet \quad (i = 0...n)$$
$$N_i \rightarrow N_{i+1}: msg_E = E < K_{prv}, plantext > \quad (4)$$
$$N_{i+1} \rightarrow N_{i+n}: msg_E = E < K_{prv}, D < K_{prv}, msg_E >>$$

If the delay by hop communication includes encryption delay, decryption delay and data transfer delay, total delay is the following:

$$T_{hop-by-hop} = t_{Enc} + t_{Transmition} + t_{Dec} + \Delta t \quad (5)$$

The $\Delta t$ in equation (5) represents the delay for the allocation and channel access. It has between zero and $T_{hop-by-hop}$. When the general node and the cluster head communicate to the encrypted packet data, the generated total delay is the following:

$$T_{tot} = \sum_{i=1}^{n} i \times (T_{hop-by-hop}) \quad (1 < n) \quad (6)$$

The n in equation (6) represents the total hop counts. It has more than 1 for the communication by the neighbor node.

Figure 12 shows that total delay according to the count of hop between *CH* and $N_i$. We assume that the encryption delay is 449*ms*, decryption delay is 456*ms* and data transfer delay is 10 ms in 16 byte data. And the number of nodes in the entire network is 215 which is less than the maximum number of nodes 65,535 in the WPAN area. We does not consider the channel access and allocation delay.

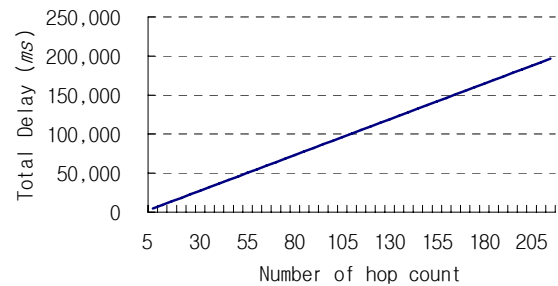

Figure 12. Total delay according as the count of hop

In figure 5, the delay of 30 hops and 180 hops generate 27,450*ms*, 164,700*ms* respectively. If the number of nodes in the entire network is 65,535 (the maximum number of nodes in the sensor network [1]), the delay is measured 59,964,525*ms* (about 16 hours). The fundamental reason of the extensive delay occurred is the performance of the equipment that used in the experiment as 8-bit Microcontroller has a low capability of the operation. Therefore, the scale of sensor network consisted of the equipments increases, the transmission delay and energy consumption will also increases.

## V. CONCLUSIONS

In this paper, we analyse the performance of AES encryption algorithm in the symmetric key encryption on ATmega644p in 8-bit microcontroller. In application scenario, we measure the encryption and decryption operation time by the plantext size. As a result, scale of the sensor network grows, the delay has been doubled. And energy consumption has also increased accordingly. In the future, specific researching on the performance analysis under plantext size and hop count require.

### ACKNOWLEDGMENT

This work was supported by the IT R&D program of MKE/IITA [2008-S-041-01, Development of Sensor Network PHY/MAC for the u-City]

AUTHORS PROFILE

H. Lee. Author is with the Department of Computing, M.Sc. course, Soongsil University, Seoul, Korea. His current research interests focus on the communications in wireless sensor networks (e-mail:hglee@cherry.ssu.ac.kr).

K. Lee. Author is with the Department of Computing, Ph.D. course, Soongsil University, Seoul, Korea. Her current research interests focus on the communications in wireless sensor networks (e-mail:khlee@cherry.ssu.ac.kr).

Y. Shin. Author was with the Computer Science Department M.Sc. and Ph.D., University of Iowa. He is now with the Professor, Department of Computing, Soongsil University. (e-mail: shin@ssu.ac.kr).